\documentclass[prb,twocolumn,superscriptaddress,showpacs,amsmath,amssymb]{revtex4-1}

\usepackage{graphicx}
\usepackage{dcolumn}
\usepackage{bm}
\usepackage{epstopdf}
\usepackage{upgreek}

\newcommand{\bra}[1]{\langle #1|}
\newcommand{\ket}[1]{|#1\rangle}

\begin{document}

\title{Anisotropic type-I superconductivity and anomalous superfluid density in OsB$_2$}

\author{J. Bekaert}
\email{jonas.bekaert@uantwerpen.be}
\affiliation{%
 Condensed Matter Theory group,
 Department of Physics, University of Antwerp,
 Groenenborgerlaan 171, B-2020 Antwerp, Belgium
}%
 \author{S. Vercauteren}%
 \affiliation{%
 Condensed Matter Theory group,
 Department of Physics, University of Antwerp,
 Groenenborgerlaan 171, B-2020 Antwerp, Belgium
}%
 \author{A. Aperis}%
 \affiliation{%
 Department of Physics and Astronomy, 
 Uppsala University, PO Box 516, 
 SE-75120 Uppsala, Sweden
}%
 \author{L. Komendov\'{a}}
 \affiliation{%
 Condensed Matter Theory group,
 Department of Physics, University of Antwerp,
 Groenenborgerlaan 171, B-2020 Antwerp, Belgium
}%
 \affiliation{%
  Department of Physics and Astronomy, 
 Uppsala University,  
 SE-75120 Uppsala, Sweden
}%
 \author{R. Prozorov}
 \affiliation{%
Department of Physics and Astronomy \& Ames Laboratory,
 Iowa State University,
 Ames, Iowa 50011, United States of America
}%
 \author{B. Partoens}
 \affiliation{%
 Condensed Matter Theory group,
 Department of Physics, University of Antwerp,
 Groenenborgerlaan 171, B-2020 Antwerp, Belgium
}%
 \author{M. V. Milo\v{s}evi\'{c}}
\email{milorad.milosevic@uantwerpen.be}
 \affiliation{%
 Condensed Matter Theory group,
 Department of Physics, University of Antwerp,
 Groenenborgerlaan 171, B-2020 Antwerp, Belgium
}%

\date{\today}

\begin{abstract}
\noindent We present a microscopic study of superconductivity in OsB$_2$, and discuss the origin and characteristic length scales of the superconducting state. From first-principles we show that OsB$_2$ is characterized by three different Fermi sheets, and we prove that this fermiology complies with recent quantum-oscillation experiments. Using the found microscopic properties, and experimental data from the literature, we employ Ginzburg-Landau relations to reveal that OsB$_2$ is a distinctly type-I superconductor with very low Ginzburg-Landau parameter $\kappa$ -- a rare property among compound materials. We show that the found coherence length and penetration depth corroborate the measured thermodynamic critical field. Moreover, our calculation of the superconducting gap structure using anisotropic Eliashberg theory and ab-initio calculated electron-phonon interaction as input reveals a single but anisotropic gap. The calculated gap spectrum is shown to give an excellent account for the unconventional behavior of the superfluid density of OsB$_2$ measured in experiments as a function of temperature. This reveals that gap anisotropy can explain such behavior, observed in several compounds, which was previously attributed solely to a two-gap nature of superconductivity. 
\end{abstract}

\maketitle

%

\section{Introduction}

\begin{figure*}[ht!!!!!!]
\centering
\includegraphics[width=0.9 \linewidth]{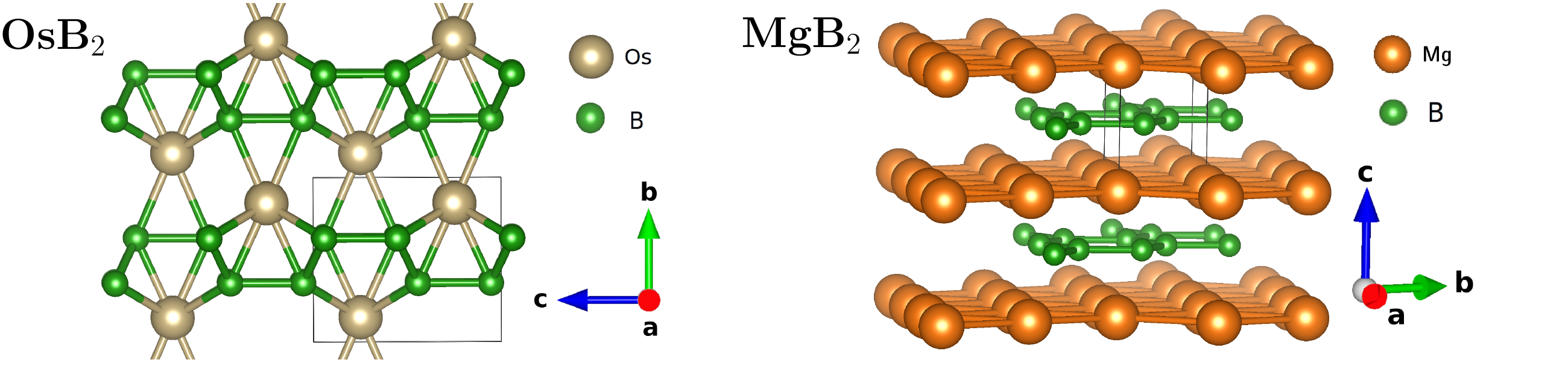}
\caption{(Color online) (a) The orthorhombic crystal structure of OsB$_2$. (b) The hexagonal crystal structure of MgB$_2$. In both cases, the unit cells are indicated by grey boxes.} 
\label{fig:fig1}
\end{figure*}

The question of particular types of superconductivity emerged in the early years of the Ginzburg-Landau (GL) theory \cite{tagkey1965a, Abrikosovb}. In type-I superconductors under applied magnetic field, an interface between normal (N) and superconducting (S) domains is energetically costly, causing normal domains to merge, whereas type-II superconductors minimize the normal domains to single vortices (which repel and organize in an Abrikosov lattice \cite{Abrikosovb}). Type-I superconductivity mainly occurs in elemental metals (Pb, Sn, In, Al, etc.), but is very rare in compounds. The fact that virtually all superconducting compounds discovered since the early 1960s are type-II superconductors \cite{Abrikosovc} (with a few notable exceptions such as YbSb$_2$, TaSi$_2$, etc.~\cite{[{}] [{ and references therein.}]PhysRevB.72.180504,PhysRevB.85.214526}) reduced the interest in type-I superconductors, until modern experimental and numerical techniques enabled more careful investigations of their rich intermediate state due to demagnetization effects in applied magnetic field: topological hysteresis of tubular/laminar domains depending on sample shape \cite{PhysRevLett.98.257001,PhysRevB.72.020505,PhysRevLett.103.267002}, the `suprafroth' ground-state \cite{Prozorov2008}, mesoscopic effects \cite{PhysRevB.83.224504,PhysRevLett.109.197003}, intricate dynamics of normal domains under applied current \cite{PhysRevB.85.092502}, etc. Moreover, a transitional regime between standard types I and II exists, that has been noted in both experiment \cite{exp_t21,PhysRevB.7.136} and microscopic theory \cite{PhysRevLett.26.629,PhysRevB.4.3029,PhysRevB.93.174503}. Its rich physics can become accessible by adding a controlled rate of non-magnetic impurities in a type-I superconductor \cite{PhysRevB.7.136,PhysRevB.90.180502}.\\
\indent One recently studied superconducting compound of which the length scales remained elusive is osmium diboride (OsB$_2$). Its rather low critical temperature (2.1 K) and the recently measured magnetization and heat-capacity of single-crystal samples \cite{PhysRevB.82.144532} pointed at possible type-I superconducting behavior, but that was not corroborated by simplified estimates of the superconducting length scales and the Ginzburg-Landau parameter $\kappa \sim 1-3 \gg 1/\sqrt{2}$. OsB$_2$ displayed additional unconventional properties, notably the temperature dependence of the superfluid density that deviates from the Bardeen-Cooper-Schrieffer (BCS) result.  In order to explain this observation, a two-gap model was proposed for superconductivity in OsB$_2$ \cite{PhysRevB.82.144532}. Although two-gap superconductivity was predicted theoretically already in 1959 \cite{PhysRevLett.3.552}, it was first identified unequivocally in MgB$_2$ only in 2001-2002 \cite{Nagamatsu2001,Choi2002}. Other materials that are candidates for multi-gap superconductors are NbSe$_2$ \cite{PhysRevLett.90.117003,PhysRevB.82.014524} and FeSe \cite{PhysRevB.80.024518,1742-6596-568-2-022005}. A successful extension of the Bardeen-Cooper-Schrieffer (BCS) theory for conventional superconductivity, with which the gap structure of superconductors can be studied, is (Migdal-)Eliashberg theory \cite{Grimvall}. In this theory, the electron-phonon interaction is not assumed instantaneous, but is retarded on the time scales determined by the phonons. In its most general form, the full reciprocal space dependence of the electron-phonon coupling is taken into account; we will refer to this treatment as fully anisotropic Eliashberg theory from here on. The technique is particularly important for identifying and studying multigap as well as anisotropic superconductors. One of the greatest successes of Eliashberg theory is the definitive proof of multigap superconductivity in MgB$_2$, from first principles, as delivered first by Choi et al. \cite{Choi2002}, and confirmed independently several times thereafter \cite{PhysRevB.87.024505,PhysRevB.92.054516}.\\
\indent Here, we present an extensive theoretical study that clarifies all of the anomalous properties of OsB$_2$ outlined above. Based on a combination of first-principles calculations and mean-field theory, we provide proof of deeply type-I behavior in OsB$_2$. Detailed knowledge of microscopic parameters and superconducting length scales obtained in this study enabled us to perform a very accurate analysis of the experimental data of Ref.~\citenum{PhysRevB.82.144532}, notably the critical magnetic field, supporting this conclusion. 
Furthermore, we reveal, based on Eliashberg calculations, that the superconducting gap spectrum of OsB$_2$ is anisotropic rather than multigap as previously proposed \cite{PhysRevB.82.144532}. These revisions of both the superconducting spectra and the length scales of OsB$_2$, starting from first principles, make an exemplary case for the interaction between experiment and theory in the field of nanostructured superconductivity.\\
\indent The paper is organized as follows: first, we discuss the crystal structure and ground state electronic structure of OsB$_2$ in Sections II and III, thoroughly making the comparison with available experimental data such as Shubnikov-de Haas measurements. We proceed by presenting in Section IV all properties related to superconductivity, namely the phonon structure, electron-phonon coupling and the gap spectrum, for which we show an excellent comparison with superfluid density measurements in Section V. Finally, in Section VI we derive the length scales of OsB$_2$ from the calculated microscopic properties using Ginzburg-Landau relations, and the resulting interaction with applied magnetic fields. Throughout, we make the comparison between the OsB$_2$ and MgB$_2$, the archetypical two-gap superconductor, pointing out both similarities and differences. Section VII summarizes our findings and conclusions.

\section{Crystal structure}

OsB$_2$ adopts the orthorhombic space group $Pmmn$ (No.~59) \cite{Int_tables}, depicted in Fig.~\ref{fig:fig1}(a). One should note a very good agreement between calculated and experimental \cite{PhysRevB.82.144532} lattice parameters, displayed in Table \ref{tab:1}, with relative deviations below 1$\%$. Os occupies Wyckoff position $2a$ depending on one internal parameter $z_{\mathrm{Os}}$ and B Wyckoff position $4f$ depending on internal parameters $x_{\mathrm{B}}$ and $z_{\mathrm{B}}$, giving a total of 6 atoms in the OsB$_2$ unit cell. The internal parameters compare equally well with experimental values (added between parentheses): $z_{\mathrm{Os}}=0.155$ (0.153), $x_{\mathrm{B}}=0.056$ (0.049) and $z_{\mathrm{B}}=0.638$ (0.641). For comparison, we show in Fig.~\ref{fig:fig1}(b) the crystal structure of MgB$_2$ (hexagonal space group P6/mmm), that is clearly layered in consecutive planes of Mg and B, as opposed to the structure of OsB$_2$. 
\begin{table}[h]
\centering
\begin{tabular}{|c|c|c|c|}
\hline
Parameter & Calc. (\AA) & Exp. (\AA) \cite{PhysRevB.82.144532}& relative dev. ($\%$)\\ \hline \hline
$a$&2.893&2.870&$+0.8$\\ \hline
$b$&4.098&4.079&$+0.5$\\ \hline
$c$&4.705&4.673&$+0.7$\\ \hline
\end{tabular}
\caption{Lattice parameters of OsB$_2$: a comparison between calculations and experiment \cite{PhysRevB.82.144532}, including the relative deviation between them.}
\label{tab:1}
\end{table}

\section{Electronic properties and Shubnikov-de Haas measurements}

\begin{figure}[t!!!!]
\centering
\includegraphics[width=\linewidth]{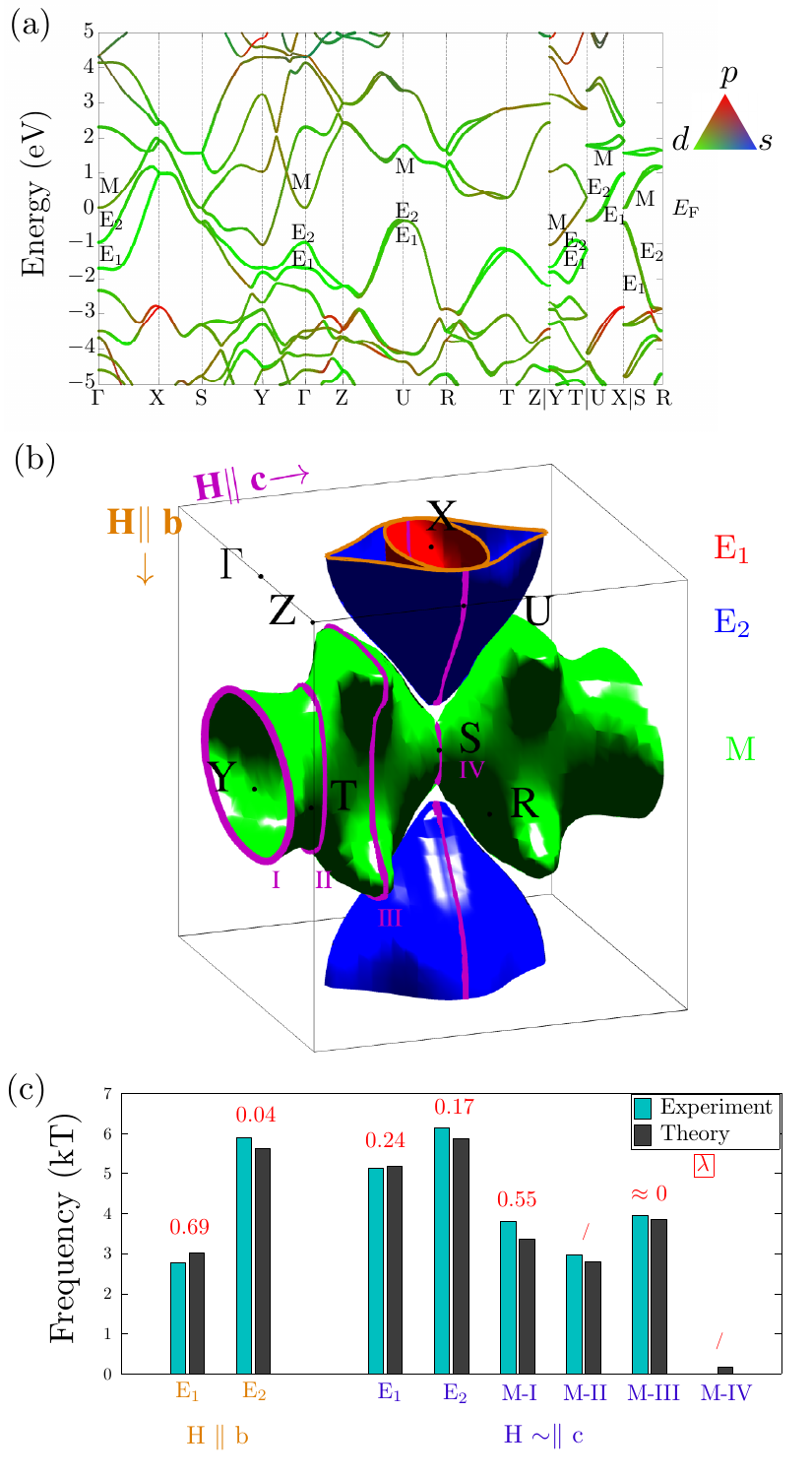}
\caption{(Color online) (a) The calculated band structure of OsB$_2$ around the Fermi level $E_{\mathrm{F}}$. The color code denotes the character of the band ($s$, $p$ or $d$), while the line thickness denotes the band character varying between pure Os (thickest) and pure B (thinnest). (b) The Fermi surface, consisting of 3 sheets: two quasi-ellipsoidal sheets E$_1$ (red) and E$_2$ (blue) and an anisotropic sheet M (green). Shubnikov-de Haas orbits in applied magnetic fields $\textbf{H} \parallel \textbf{b}$ and $\textbf{H} \parallel \textbf{c}$ are also indicated, $ \textbf{b}$ and $ \textbf{c}$ being unit cell vectors. (c) A comparison between our theoretical calculations for Shubnikov-de Haas frequencies (in units of kilotesla -- kT) and the experimental values from Ref. \citenum{PhysRevB.82.144532}. The resulting electron-phonon coupling values $\lambda$ are added in red. Slashes indicate that the experimental cyclotron masses were not available.} 
\label{fig:fig2}
\end{figure}

We start from a first-principles study of the electronic structure of OsB$_2$ based on density functional theory (DFT), implemented in VASP \cite{Kresse}. In this study -- for which computational details can be found in Appendix A -- we take into account spin-orbit coupling, in view of the high atomic number of Os. The band structure according to orbital character, shown in Fig.~\ref{fig:fig2}(a), reveals predominant Os-$d$ character of the bands crossing the Fermi level ($E_{\mathrm{F}}$). A fraction of B-$p$ states also contributes to the band we denote M because of this mixed character. A total of three bands is present at $E_{\mathrm{F}}$, so the resulting Fermi surface, depicted in Fig.~\ref{fig:fig2}(b), consists of three sheets. First, there are two nested quasi-ellipsoidal sheets with pure Os-$d$ character, centered around X, the inner one denoted E$_1$ and the outer one E$_2$. The third sheet M, with central axis along direction Y-S, is more anisotropic. One of the most successful experimental techniques to probe Fermi surfaces is the one of quantum oscillations, utilizing the Shubnikov-de Haas (SdH) effect. In this effect, the conductivity of a metal shows oscillations with frequencies proportional to the areas of extremal orbits of the Fermi surface $A(E_{\mathrm{F}})$, perpendicular to the applied magnetic field \cite{Abrikosov}. The amplitude of the SdH oscillations depends on the cyclotron mass of the electrons dressed with phonon interaction $m_{\mathrm{c}}^*$. The extremal orbits in the case of OsB$_2$ are indicated in Fig.~\ref{fig:fig2}(b) for two different magnetic fields. We calculated the SdH frequencies $f$ and bare cyclotron masses $m_{\mathrm{c}}$ (i.e. \textit{without} phonon dressing), respectively as
 $f=\frac{\hbar}{2 \pi e}A(E_{\mathrm{F}})$ and $m_{\mathrm{c}}=\frac{\hbar^2}{2 \pi}\left.\left(\frac{\partial A}{\partial E}\right)\right|_{E=E_{\mathrm{F}}}$,
where the derivative w.r.t.~$E$ is treated within central difference approximation. In our simulation we account for the fact that the applied field was not exactly parallel to unit cell vector $\textbf{c}$ in the corresponding experiment (noted in Ref.~\citenum{PhysRevB.82.144532}). The agreement between theoretical and experimental SdH frequencies, displayed in Fig.~\ref{fig:fig2}(c), is excellent (for field angle 23$^\circ$ with respect to $\textbf{c}$ in the \textbf{ac} plane, in the simulations), validating the electronic structure of OsB$_2$ presented in Figs.~\ref{fig:fig2} (a) and (b).  Through the mass renormalization relation \cite{Grimvall} $m_{\mathrm{c}}^*=(1+\lambda)m_{\mathrm{c}}$, the electron-phonon (\textit{e-ph}) coupling $\lambda$ in specific orbits can also be estimated. This reveals that the \textit{e-ph} coupling in E$_1$ and E$_2$ is strongly anisotropic. The coupling is strong in E$_1$ in the orbit perpendicular to $\textbf{b}$, while it is very small in E$_2$. The orbits of E$_1$ and E$_2$ perpendicular to $\textbf{c}$ show similar, moderate \textit{e-ph} coupling. Band M has no closed orbits perpendicular to $\textbf{b}$, but the \textit{e-ph} coupling in the other direction ranges from almost zero to quite strong, depending on the orbit. The most important conclusion from this analysis is that \textit{all} bands contribute to the \textit{e-ph} coupling. The question of how the superconducting gap is distributed over the bands will be treated in the following sections.

\section{Phonons and electron-phonon interaction}
\begin{figure*}[t]
\centering
\includegraphics[width=0.9 \linewidth]{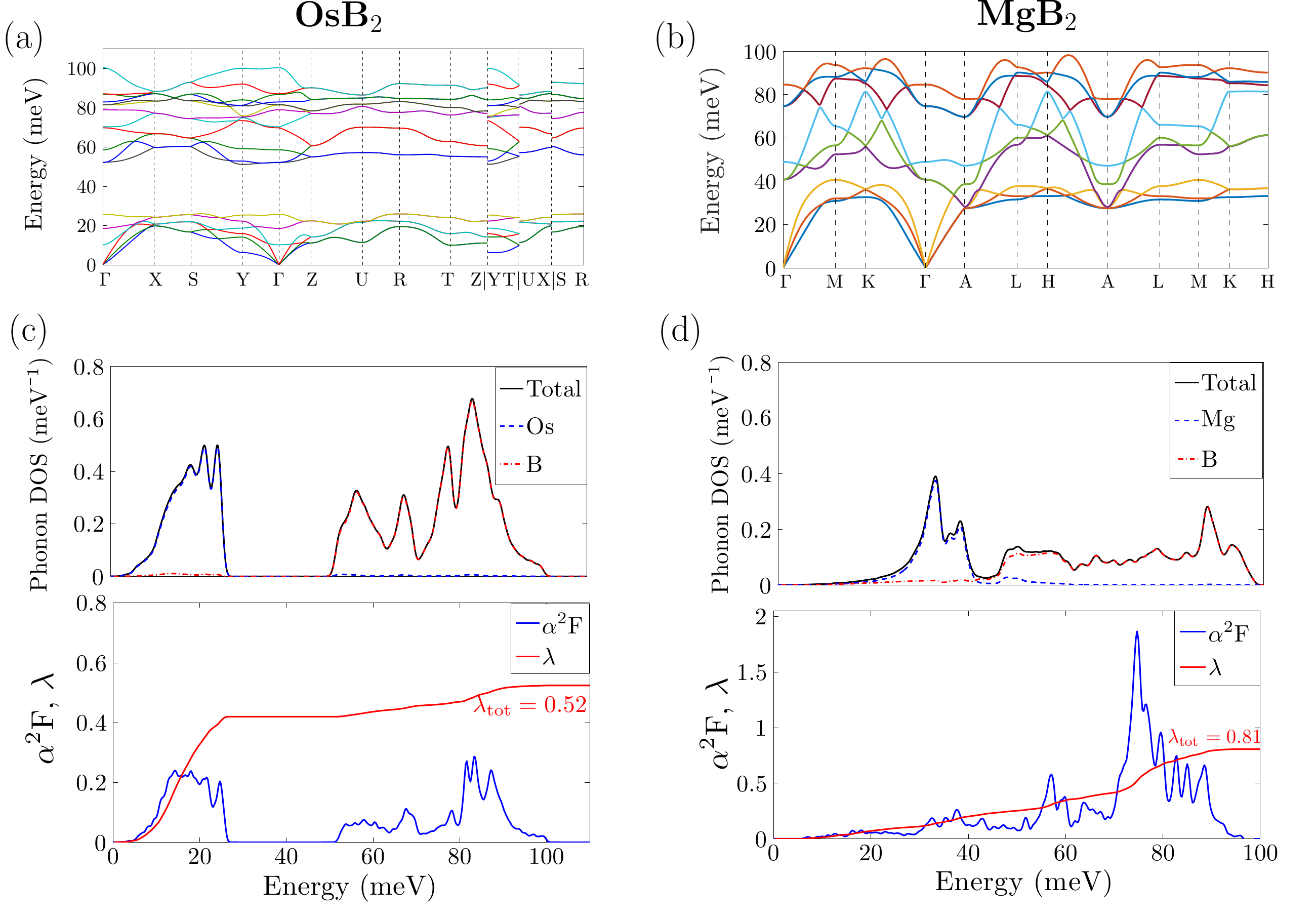}
\caption{(Color online) (a) and (b) The phonon band structure calculated using density-functional perturbation theory, for OsB$_2$ and MgB$_2$ respectively. (c) and (d) Top panel: The phonon DOS of OsB$_2$ and MgB$_2$ respectively, split into contributions of Os/Mg and B. Bottom panel: The Eliashberg function $\alpha^2$F describing the energy-dependent electron-phonon coupling, and the resulting isotropic electron-phonon coupling constant $\lambda_{\mathrm{tot}}$.}
\label{fig:fig3}
\end{figure*}
\begin{figure*}[t]
\centering
\includegraphics[width=0.8 \linewidth]{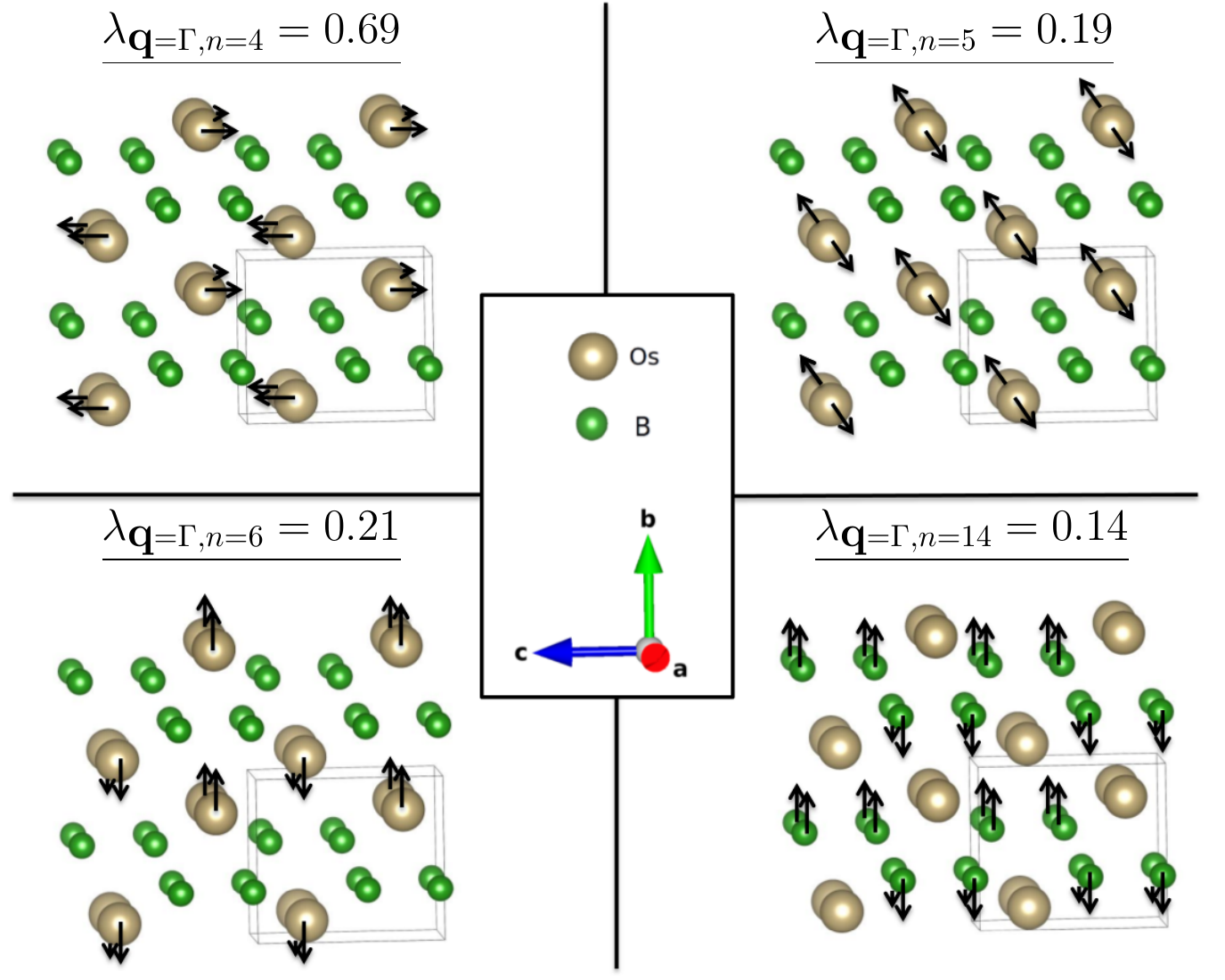}
\caption{(Color online) Atomic displacements of phonon modes that couple strongly to electrons in OsB$_2$.  In each case $\lambda_{\textbf{q},n}$ is given. Modes $n=4,5,6$ are optical modes of Os, along the three crystal axes. Mode $n=14$, with displacements along axis \textbf{b}, is the optical B-mode with strongest coupling to electrons. The inset shows the legend for the atom types and the crystal axes.}
\label{fig:fig4}
\end{figure*}
In order to analyze the mechanism leading to superconductivity in OsB$_2$, we carried out a first-principles calculation of the \textit{e-ph} interaction. To this end we used Eliashberg theory combined with density functional perturbation theory (DFPT), as implemented in ABINIT \cite{PhysRevB.54.16487,Gonze20092582}, and details on which are specified in Appendix A. The phonon band structure, shown in Fig.~\ref{fig:fig3}(a) is characterized by a distinct gap of $\sim$ 25 meV. The characteristic Debye temperature is obtained from the speed of sound $v_{\mathrm{s}}$ in the material:
\begin{equation}
\Theta_{\mathrm{D}}=\frac{hv_{\mathrm{s}}}{2k_{\mathrm{B}}}\sqrt[3]{\frac{6}{\pi}N(E_{\mathrm{F}})}~,
\end{equation}
where $N(E_{\mathrm{F}})$ is the total density of states at the Fermi level. The effective speed of sound is calculated as the following average of the values due to the transversal and longitudinal acoustic modes, $v_{\mathrm{t}}$ and $v_{\mathrm{l}}$ (the slope of the phonon dispersions near $\Gamma$) \cite{Long2013336}:
\begin{equation}
v_{\mathrm{s}}=\sqrt[3]{3}\left(\frac{2}{v_{\mathrm{t}}^3}  + \frac{1}{v_{\mathrm{l}}^3}\right)^{-\frac{1}{3}}~.
\end{equation}
We find $\Theta_{\mathrm{D}}=471$ K, in good accordance with the experimental value of 550 K \cite{PhysRevB.80.014507}. The origin of the gap in the phonon band structure becomes clear in the phonon density of states (DOS) shown in Fig.~\ref{fig:fig3}(c). Owing to the high mass of Os, its phonons are low-energy ones, in contrast with B-related phonons extending up to energies of $\sim$ 100 meV. The Eliashberg function and \textit{e-ph} coupling shown in Fig.~\ref{fig:fig3}(c) point at the dominance of Os-related phonons in the coupling constant. Therefore, both the fermionic and the phononic features of superconductivity in OsB$_2$ are driven by Os. The total isotropic coupling amounts to $\lambda_{\mathrm{tot}}=0.52$. Using the McMillan-Dynes formula \cite{PhysRevB.12.905}, critical temperature $T_{\mathrm{c}}=2.1$ K is found (taking as Coulomb pseudopotential $\mu^*=0.13$).\\
\indent A comparison of the electron-phonon interaction in OsB$_2$ to the case of the well-known two-gap superconductor MgB$_2$ is instructive. 
As shown in Fig.~\ref{fig:fig3}(b), the contributions of Mg and B to the phonon DOS are again quite distinguishable, due to the mass difference, similar to the case of OsB$_2$. In the latter, the Os modes are even lower in energy because of the high atomic number of Os. From the Eliashberg function, shown in Fig.~\ref{fig:fig3}(d), we find coupling constant $\lambda_{\mathrm{tot}}=0.81$ for MgB$_2$, compared to $\lambda_{\mathrm{tot}}=0.52$ for OsB$_2$. This yields $T_{\mathrm{c}} = 24$ K for MgB$_2$, significantly higher than for OsB$_2$.  
The main contribution to this strong coupling in MgB$_2$ is the pronounced peak in the Eliashberg function around $\sim 75$ meV, stemming from B-related phonon modes, in contrast with what we established for OsB$_2$. $T_{\mathrm{c}} = 24$ K is still an underestimation of the experimental $T_{\mathrm{c}} = 39$ K for bulk MgB$_2$, due to the limitations of the isotropic Eliashberg theory. 
In reality, the electron-phonon coupling is very anisotropic in MgB$_2$ and this anisotropy has a pronounced effect on $T_{\mathrm{c}}$ \cite{PhysRevB.66.020513}. This fact has been established by combined anisotropic Eliashberg theory and DFT calculations
that have been very successful in explaining superconductivity in this material \cite{Choi2002,PhysRevB.87.024505,PhysRevB.92.054516} and also made predictions for
further experiments \cite{PhysRevB.92.054516}. The anisotropic electron-phonon coupling and the particular Fermi surface of MgB$_2$ result in two distinct superconducting gaps over different Fermi surface sheets in this material. Therefore, it is possible to obtain an effective isotropic two band model that captures the essential characteristics of two-gap superconductivity in MgB$_2$ \cite{PhysRevB.73.104520}. In this case, the coupling is described by a $2 \times 2$ matrix of coupling constants. For MgB$_2$, it has been measured to be \cite{Kuzmichev2014}
$\Lambda=\begin{pmatrix} 0.84 & 0.19 \\ 0.19 & 0.39 \end{pmatrix}$, with the largest eigenvalue of this matrix playing the role of an effective coupling constant in the multigap case \cite{PhysRevLett.87.087005}: $\lambda_{\mathrm{eff}}=0.91$. In this approach, the multigap effect accounts for a higher $T_{\mathrm{c}}=37$ K (using $\mu^*=0.1$). As we show in the next section, the application of a similar effective two-gap model to OsB$_2$ leads to incorrect conclusions about the nature of the superconducting state of the material.\\
\indent In MgB$_2$, the dominant phonon mode in the \textit{e-ph} coupling is the in-plane hexagon deformation mode E$_{2g}$ of the B atoms \cite{Choi2002}. In OsB$_2$, on the other hand, 80\% of all \textit{e-ph} coupling is contributed by Os-related modes. The strongest coupling resides in the three optical modes of Os, with energy values between 9 and 26 meV, cf.~Fig.~\ref{fig:fig3}. Although spread over \textbf{q}-space \footnote{For the definition of the \textbf{q}-point grid, see the Appendix.}, the coupling in these modes is strongest at $\textbf{q}=(0,0,0)=\Gamma$, thus promoting intraband coupling. In its turn, it bears important consequences for the superconducting gap spectrum, as we will show in the next section. The atomic displacements corresponding to the different optical modes of Os (with mode numbers $n=4,5,6$) at $\Gamma$ are shown in Fig.~\ref{fig:fig4}. The displacements are directed along the three crystal axes, along \textbf{c}, \textbf{a} and \textbf{b} for $n=4,5,6$ respectively. The mode with the lowest energy (the softest mode), $n=4$,  carries the strongest \textit{e-ph} coupling $\lambda_{\textbf{q}=\Gamma,n=4}=0.69$, compared to $\lambda_{\textbf{q}=\Gamma,n=5}=0.19$ and $\lambda_{\textbf{q}=\Gamma,n=6}=0.21$ for the other two modes. The residual 20\% of the total \textit{e-ph} coupling is contributed by B-related optical modes. It is strongest in mode $n=14$, at $\Gamma$, and corresponds to a displacement of the B-atoms along \textbf{b}, as shown in Fig.~\ref{fig:fig4}, and leads to the peak in $\alpha^2F$ at 81 meV.  

\section{Anisotropic superconducting gap and anomalous superfluid density}
\begin{figure}[t!!!!]
\centering
\includegraphics[width=\linewidth]{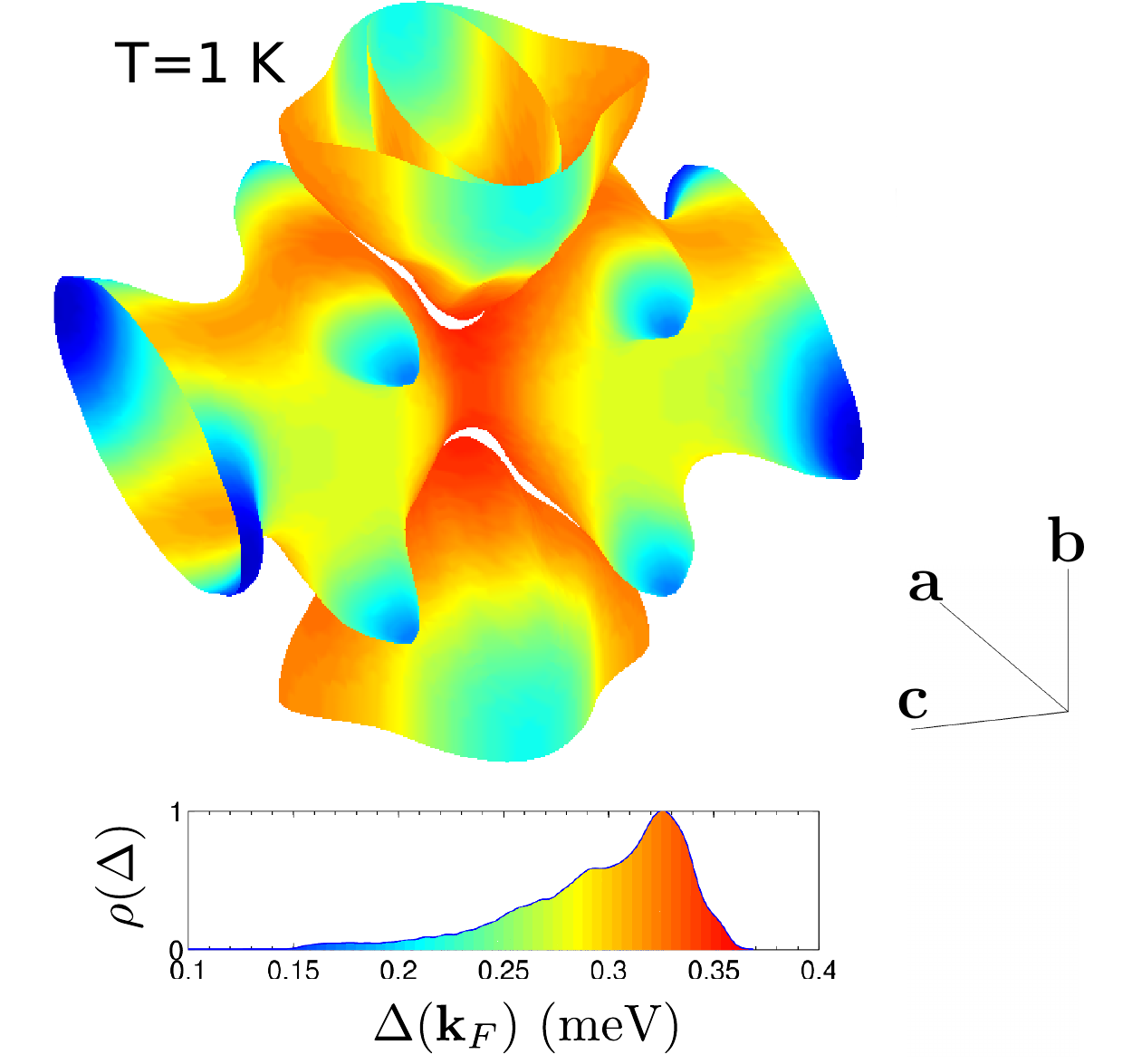}
\caption{(Color online) The superconducting gap spectrum of OsB$_2$ on the Fermi surface, calculated using fully anisotropic Eliashberg theory at $T=1$ K, using the electron-phonon coupling obtained from first-principles as input. $\rho(\Delta)$ is the distribution of the gap, thus showing a single anisotropic gap.} 
\label{fig:fig5}
\end{figure}
\begin{figure}[t!!!!]
\centering
\includegraphics[width=\linewidth]{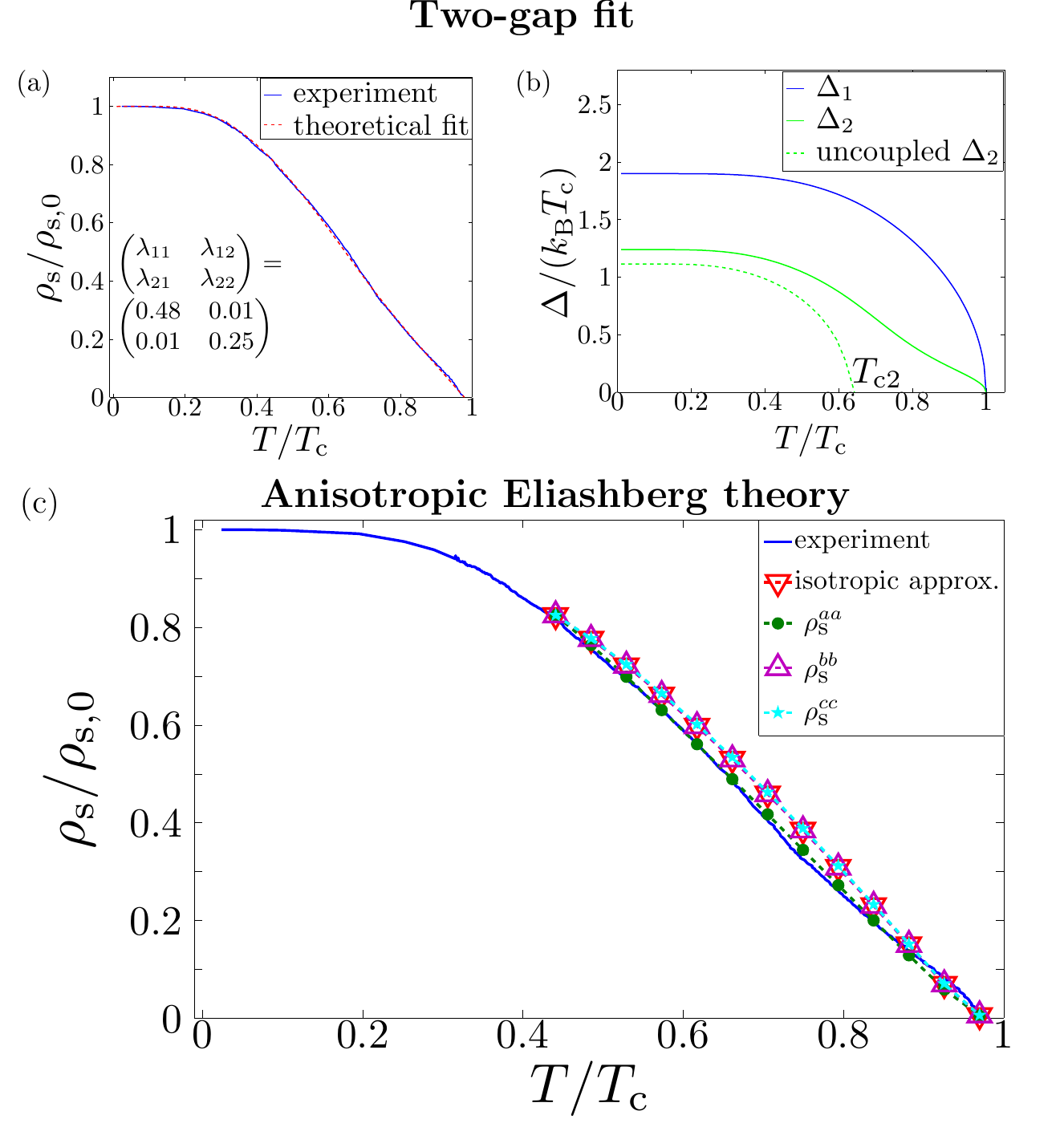}
\caption{(Color online) (a) The two-gap $\gamma$ fit of the superfluid density $\rho_{\mathrm{s}}$ from Ref.~\citenum{PhysRevB.82.144532} (normalized with the superfluid density at zero temperature $\rho_{\mathrm{s},0}$), using the microscopic parameters calculated here from first-principles. The matrix of coupling constants obtained from the fit is shown as inset. (b) The gap profiles as a function of temperature, obtained by solving the BCS gap equations for two coupled condensates. Dashed line shows the weaker gap ($\Delta_2$) in the absence of interband coupling, with $T_{\mathrm{c2}}=1.35$ K. (c) The superfluid density calculated from anisotropic Eliashberg theory, both isotropic and along specific directions. Component $\rho^{aa}_{\mathrm{S}}$ matches the experimental data extremely well.} 
\label{fig:fig6}
\end{figure}
Starting from the electron-phonon interaction obtained in the previous section, we can calculate the superconducting gap spectrum using fully anisotropic Eliashberg theory. Technical aspects of this calculation can be found in Appendix A and in Ref.~\citenum{PhysRevB.92.054516}. The resulting superconducting gap spectrum $\Delta(\textbf{k})$ of OsB$_2$, at an intermediate temperature of $T=1$ K, is displayed in Fig.~\ref{fig:fig5}. The distribution $\rho(\Delta)$ shows that the gap varies continuously over all Fermi sheets. Therefore, OsB$_2$ is identified as an \textit{anisotropic}, due to the spread of the gap spectrum, but \textit{single-gap superconductor}. It is apparent that the gap spectrum is entirely symmetric within the \textbf{bc}-planes, but has a strong evolution along the \textbf{a}-axis. We find a rather strong electron-electron interaction to be at play between the condensed electrons, since a high Coulomb pseudopotential, $\mu^{*}=0.215$, is needed to get the correct $T_{\mathrm{c}}$ in the Eliashberg calculations. As a result of this and the moderate electron-phonon coupling, the gap values are on the low side, ranging between 0.15 and 0.37 meV at 1 K.\\
\indent This result of a single, anisotropic gap in OsB$_2$ seems to contradict the findings in Ref.~\citenum{PhysRevB.82.144532} at first sight, where two-gap superconductivity in OsB2 was suggested based on a successful fit of the two-gap $\gamma$ model \cite{PhysRevB.80.014507}. We show in Fig.~\ref{fig:fig6} an updated version of this fit, using the calculated density of states per band. The obtained coupling constant matrix, shown as inset in Fig.~\ref{fig:fig6}(a), is subsequently used in Bardeen-Cooper-Schrieffer (BCS) gap equations to calculate the evolution of both gaps with temperature, displayed in Fig.~\ref{fig:fig6}(b). The obtained values of the gaps at zero temperature are $\Delta_1(0)=0.36$ meV and $\Delta_2(0)=0.24$ meV. It follows thus that a two-gap superconductivity model is \textit{sufficient} to fit the superfluid density measurements, but is it \textit{necessary}? To answer this question, we calculate the superfluid density within anisotropic Eliashberg theory. The normalized superfluid density tensor is then given by
\begin{equation}
\frac{\rho^{\alpha\beta}_s(T)}{\rho^{\alpha\beta}_s(0)}= T\sum_{\omega_n}  \langle (\nabla_\alpha E_{{\bf k}}\nabla_\beta E_{{\bf k}})\frac{\Delta^2_{{\bf k},n}}{Z_{{\bf k},n}\left[\omega^2_n +\Delta^2_{{\bf k},n}\right]^{\frac{3}{2}}}\rangle_{k_\mathrm{F}}~,
\label{eq:superfl_den}
\end{equation}
where $\omega_n$ are the Matsubara frequencies, $E_{\textbf{k}}$ signifies the normal-state electronic spectrum, $Z_{\textbf{k},n}$ is the mass renormalization in Eliashberg theory \cite{Grimvall} and $\langle ... \rangle_{k_{\mathrm{F}}}$ denotes the Fermi surface average. 
We plot the evolution of the different elements of the superfluid density tensor as a function of temperature in Fig.~\ref{fig:fig6}(c). For OsB$_2$, all off-diagonal terms of the superfluid density tensor are zero. In the isotropic approximation, $\nabla_\alpha E_{{\bf k}}\nabla_\beta E_{{\bf k}}$ (product of Fermi velocity components) is pulled out of the Fermi surface average in Eq.~(\ref{eq:superfl_den}). Within this approximation, the superfluid density matches $\rho_{\mathrm{S}}^{bb}=\rho_{\mathrm{S}}^{cc}$. Along the \textbf{b} and \textbf{c} directions, the superfluid density is the same, due to the \textbf{bc}-symmetry of the superconducting gap spectrum that we pointed out earlier. The superfluid density along the \textbf{a} direction, however, is significantly different and matches the experimental measurement extremely well.
 In the case of OsB$_2$, the convex shape of $\rho_{\mathrm{S}}^{aa}$ is not a result of the multigap character \cite{PhysRevLett.3.552}, but follows naturally from the temperature evolution of the anisotropic condensate. This comparison of Eliashberg theory to the experiment provides a clear example of an anomalous superfluid density of a \textit{single gap} superconductor, and hence a caveat for future identifications of multigap superconductors. 

\section{Type-I behavior}
\begin{figure}[t]
\centering
\includegraphics[width=\linewidth]{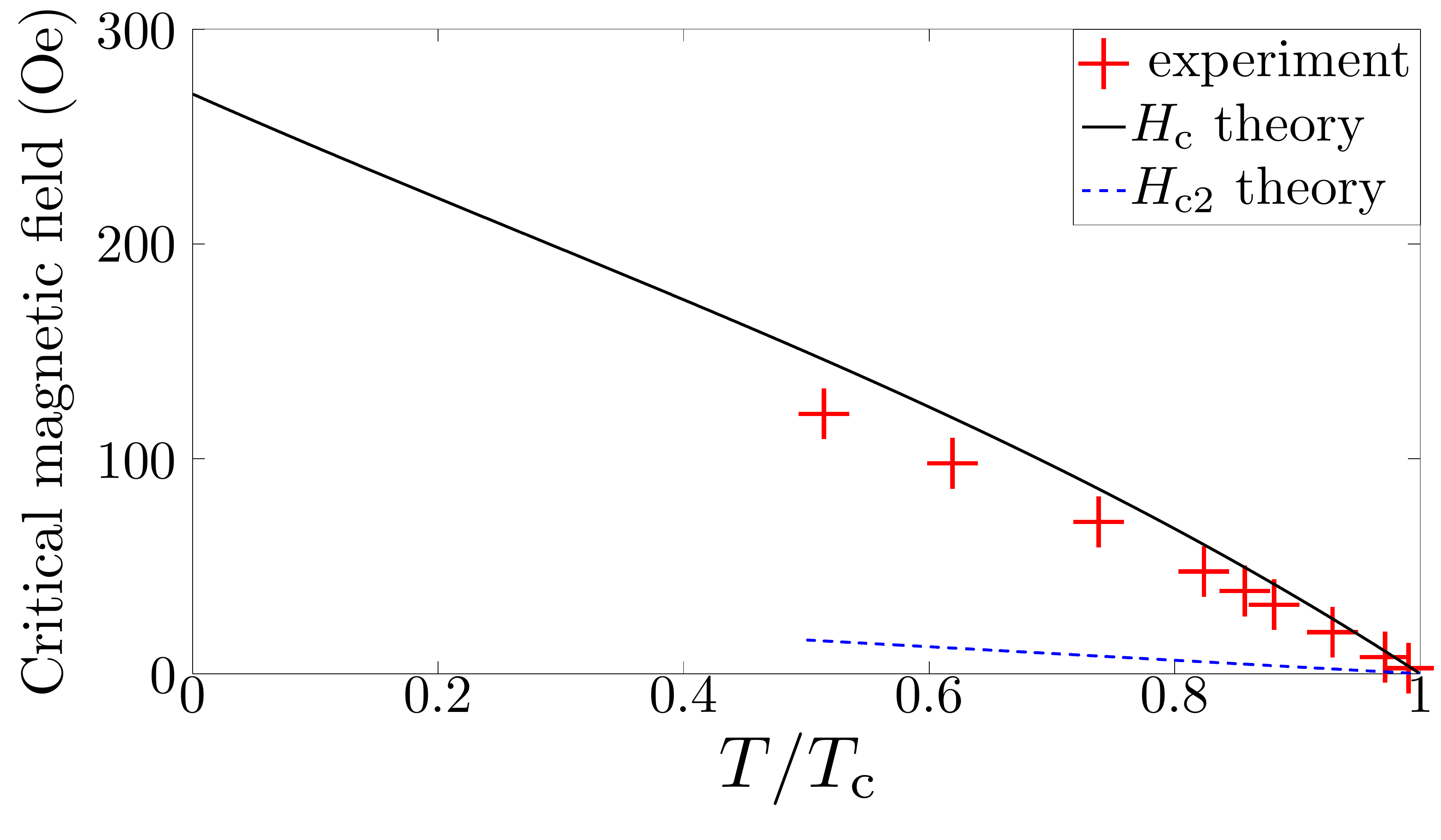}
\caption{(Color online) The thermodynamic critical magnetic field $H_{\mathrm{c}}$ calculated using the extended Ginzburg-Landau formalism \cite{PhysRevB.85.014502}, compared to experimental data from Ref.~\citenum{PhysRevB.82.144532}. To further illustrate the strong type-I character of OsB$_2$ we include the calculated upper critical field $H_{\mathrm{c}2}$ to visualize that $H_{\mathrm{c}2} \ll H_{\mathrm{c}}$.}
\label{fig:fig7}
\end{figure}
To further understand the superconducting behavior of OsB$_2$, particularly under applied magnetic field, we calculate the characteristic length scales of superconductivity, i.e. coherence length $\xi(0)=\frac{\hbar v_{\mathrm{F}}}{4\pi T_{\mathrm{c}}}\sqrt{\frac{7\zeta(3)}{3}}$, and London penetration depth $\lambda_{\mathrm{L}}(0)=\sqrt{\frac{3c^2}{16\pi e^2 v_{\mathrm{F}}^2N(E_{\mathrm{F}})}}$ \cite{PhysRevB.87.134510}. There are significant differences between the quasi-ellipsoids and sheet M with regard to the microscopic parameters. The former account for a density of states of 0.23 states/eV per formula unit, whereas the latter occupies 0.39 states/eV per formula unit. On the other hand, the quasi-ellipsoidal sheets are more highly curved than sheet M, with respective average Fermi velocities of $6.5 \cdot 10^5$ m/s and $3.7 \cdot 10^5$ m/s. 
Nevertheless, since a single condensate in OsB$_2$ was found in the previous section, we perform a weighted average over the whole Fermi surface within mean-field theory. This leads to $N(E_{\mathrm{F}})=0.62$ states/eV per formula unit and $v_{\mathrm{F}}=4.7 \cdot 10^5$ m/s. The resulting length scales are $\lambda_{\mathrm{L}}(0)=27$ nm and $\xi(0)=229$ nm. The GL parameter $\kappa=\lambda_{\mathrm{L}}/\xi=0.12$ is therefore far below $\kappa=1/\sqrt{2}$, the value separating type-I from type-II superconductors \cite{Abrikosovb}, indicating that OsB$_2$ is deeply in the type-I regime. The small penetration depth and large coherence length in OsB$_2$ occur due to the large $v_{\mathrm{F}}$, comparable to the high Fermi velocities in elemental metals, and due to the rather low $T_{\mathrm{c}}$.\\ 
\indent To further corroborate our findings, we show the compliance of our conclusions with available experimental data. Specifically, we look at the experimental critical magnetic fields \cite{PhysRevB.82.144532}, to determine whether they correspond to the thermodynamic critical field $H_{\mathrm{c}}$ or the upper critical field $H_{\mathrm{c}2}$. To calculate $H_{\mathrm{c}}$, we use the expression from the recently developed Extended Ginzburg-Landau (EGL) theory, demonstrated to be in very good accordance with BCS results even quite far from $T_{\mathrm{c}}$ \cite{PhysRevB.85.014502,PhysRevB.86.144514}. The thereby obtained $H_{\mathrm{c}}$ of OsB$_2$ is presented in Fig.~\ref{fig:fig7} -- revealing a very good agreement with the experimental values, in a broad temperature range. The non-linearity of $H_{\mathrm{c}}$ away from $T_{\mathrm{c}}$ is captured by the EGL theory. As expected for a pure type-I superconductor, the calculated upper critical field $H_{\mathrm{c}2}=\frac{\Phi_0}{4\pi \xi^2(0)}\tau$, where $\tau=1-\frac{T}{T_{\mathrm{c}}}$, is much lower, since $H_{\mathrm{c}2}/H_{\mathrm{c}}\propto \kappa$. The fact that EGL theory, in combination with the calculated microscopic parameters, predicts $H_{\mathrm{c}}$ so well, yields another proof of the type-I behavior of OsB$_2$, be it indirect. For direct experimental proof, imaging of the intermediate state of OsB$_2$ should be performed, which may show the large normal domains characteristic of type-I materials, but may also reveal surprises related to the anisotropy of the gap spectrum.

\section{Conclusion}
In summary, we presented solid and multiscale proofs for anisotropic type-I superconductivity in OsB$_2$, combining first-principles calculations, mean field theory and recent experimental data. The Fermi surface of OsB$_2$ consists of two nested quasi-ellipsoidal Fermi sheets with Os-$d$ character and a third sheet with mixed band character. From a first-principles calculation of the electron-phonon coupling, we found that OsB$_2$ has very moderate coupling amounting to the isotropic value $\lambda=0.52$. The main contribution to this value (80\%) stems from the low-energy Os-related modes. This is a very different situation from the coupling in MgB$_2$, due to the entirely different crystal structure of the two compounds, where a particular optical vibration of B-atoms couples strongly with the electrons. From the electron-phonon coupling we calculated the superconducting gap spectrum using fully anisotropic Eliashberg theory. The result is a single, anisotropic gap at odds with the available two-gap fit of the superfluid density in Ref.~\citenum{PhysRevB.82.144532}. To settle this issue, we calculated the superfluid density within Eliashberg theory, taking into account the anisotropy in the Fermi surface. We found that the superfluid density along the shortest lattice axis (\textbf{a} in Fig.~\ref{fig:fig4}) matches the experimental data with excellent accuracy. Thus, OsB$_2$ provides an instructive example of an anomalous temperature dependence of the superfluid density due to a single, anisotropic gap, that cannot be fitted within the simplest BCS model. \\
\indent The Fermi velocities in OsB$_2$ are high for a compound material, while its $T_{\mathrm{c}}$ is rather low, resulting in a very low Ginzburg-Landau parameter -- setting OsB$_2$ deeply in the type-I superconducting regime. Moreover, we showed that this complies with the available measurements of the critical magnetic field. 
The here revealed characteristics of OsB$_2$ provide a general recipe for other type-I superconducting compounds to be discovered, combining moderate electron-phonon coupling (thus low $T_{\mathrm{c}}$, and long coherence lengths), and a highly curved Fermi surface (thus high Fermi velocities, diminishing $\kappa$ with a squared dependence). Such materials will in turn provide more direct access to the scarcely studied regimes of superconductivity away from the standard type-II, especially interesting in multigap superconductors and superconductors with an anisotropic gap.

\begin{acknowledgments}
\noindent This work was supported by TOPBOF-UAntwerp, Research Foundation-Flanders (FWO) and COST action MP1201. A.E. and L.K. acknowledge funding from the Swedish Research Council (Vetenskapsr\r{a}det). L.K. also acknowledges the Wenner-Gren Foundations. The first-principles calculations were carried out on the HPC infrastructure of the University of Antwerp (CalcUA), a division of the Flemish Supercomputer Centre (VSC), supported financially by the Hercules foundation and the Flemish Government (EWI Department). The anisotropic Eliashberg calculations were performed using resources from the Swedish National Infrastructure for Computing (SNIC).
\end{acknowledgments}

\begin{appendix}
\section{Computational details}
Our density functional theory (DFT) calculations make use of the Perdew-Burke-Ernzerhof (PBE) functional, including spin-orbit interaction, implemented within a planewave basis in the VASP code \cite{Kresse}. Electron-ion interactions are treated using projector augmented wave (PAW) potentials, taking into account Os-5$p^6$6$s^2$5$d^6$ and B-2$s^2$2$p^1$ as valence electrons. The energy cutoff for the plane-wave basis is set to 500 eV, to achieve convergence of the total energy below 1 meV per atom. To obtain a very accurate description of the Fermi surface, also needed for accurate calculation of the Fermi velocities and electronic density of states per band, a very dense $40 \times 32 \times 24$ $ \Gamma$-centered Monkhorst-Pack \textbf{k}-point grid is used. For high-symmetry \textbf{k}-points, we use the notational convention established in Ref.~\cite{Curtarolo}. The optimized crystal structure was obtained using a conjugate-gradient algorithm so that forces on each atom were below 1 meV/\AA.\\
\indent Density functional perturbation theory (DFPT) calculations were carried out within the framework of ABINIT \cite{Gonze20092582}, keeping the same valence electrons as in VASP, and also using the PBE functional.  
The crystal structure was optimized again in ABINIT, with no significant differences with the values reported in Table \ref{tab:1}. The total number of perturbations due to atomic displacements (in other words, the number of phonon branches) amounts to $3\cdot N_{\mathrm{atoms}}=18$. In order to calculate the Eliashberg function 
\begin{equation*}
\alpha^2F(\omega)=N(E_{\mathrm{F}})\sum_{\textbf{k} \textbf{q} \nu}\vert g_{\textbf{k}, \textbf{k}+\textbf{q}}^{\nu}\vert^2\delta(\omega-\omega_{\textbf{q}\nu})~,
\end{equation*}
one needs the total density of states $N(E_{\mathrm{F}})$, the electron-phonon coupling coefficients $g_{\textbf{k}, \textbf{k}+\textbf{q}}^{\nu}$ and the phonon spectrum $\omega_{\textbf{q}\nu}$, the latter two obtained within DFPT. The electron-phonon coupling coefficients $g_{\textbf{k}, \textbf{k}+\textbf{q}}^{\nu}$ are proportional to the matrix elements $\bra{\textbf{k}+\textbf{q}}\delta V\ket{\textbf{k}}$, where $\delta V$ is the perturbative part of the Hamiltonian \cite{PhysRevB.54.16487}. We carried out the summation to obtain the Eliashberg function over a $21 \times 15 \times 15$ $\textbf{k}$-point grid and a $7 \times 5 \times 5$ $\textbf{q}$-point grid (a subgrid of the $\textbf{k}$-point grid). The isotropic electron-phonon coupling function is the first inverse moment of the Eliashberg function:
\begin{equation*}
\lambda(\omega)=2\int_0^\omega \mathrm{d}\omega'\omega'^{-1}\alpha^2F(\omega')~.
\end{equation*}
The electron-phonon coupling constant is $\lambda_{\mathrm{tot}}=\lambda(\omega_{\mathrm{max}})$, where $\omega_{\mathrm{max}}$ is the maximum phonon frequency. 
Moreover one defines
\begin{equation*}
\omega_{\mathrm{log}}=\mathrm{exp}\left( \frac{2}{\lambda_{\mathrm{tot}}}\int_0^\infty \mathrm{d}\omega \omega^{-1}\mathrm{ln}(\omega) \alpha^2F(\omega)\right)~,
\end{equation*}
with which ultimately the critical temperature can be calculated with the McMillan-Dynes formula (solution to the Eliashberg equations in the weak to intermediate coupling limit):
\begin{equation*}
T_{\mathrm{c}}=\frac{\hbar \omega_\mathrm{log}}{1.2 k_{\mathrm{B}}}\mathrm{exp}\left(-\frac{1.04 (1+\lambda_{\mathrm{tot}})}{\lambda_{\mathrm{tot}}-\mu^*(1+0.62\lambda_{\mathrm{tot}})}\right)~,
\end{equation*}
where $\mu^*$ is the renormalized Coulomb repulsion between Cooper pair electrons, the so-called `Coulomb pseudopotential' \cite{Grimvall}.\\
\indent The Eliashberg calculations were performed with the Uppsala Superconductivity code (UppSC). The anisotropic Eliashberg equations were solved self-consistently in Matsubara space, starting from the electron and phonon band structures and electron-phonon coupling obtained with DFPT. In this scheme, we iterated until convergence better than $10^{-3}$ on the relative gap values between each iteration step was reached. In all calculations, we employed standard $\mu^* = 0.215$ for the Coulomb pseudopotential, in order to match the experimental $T_{\mathrm{c}}$. For the sums over Matsubara frequencies a cut-off energy of up to 0.7 eV was used (total of 2592 Matsubara frequencies). In order to find the superconducting gap-edge, the converged solutions were
analytically continued to real frequencies with a Pad\'{e} approximation procedure.

\end{appendix}

\bibliography{biblio}

\end{document}